\documentclass[12pt,centertags,oneside,reqno]{amsart}
\usepackage{latexsym,theoremfile,graphicx,color} %myharv
\usepackage{latexsym,amssymb,amsfonts,rotate,enumerate}
\usepackage{amsmath}

\usepackage[latin1]{inputenc}

\usepackage{natbib}

\def\endpr{\hfill $\Box$ \vskip .3em \noindent }
\def\proof{\noindent{\sc Proof: }}

\theoremstyle{dissprop}
{\theoremstyle{dissdefi} \newtheorem{definition}{definition}[section] }

\newtheorem{proposition}[definition]{proposition}

\newtheorem{lemma}[definition]{lemma}
\newtheorem{corollary}[definition]{corollary}
\definecolor{red,green,blue}{rgb}{0.5,0.85,0.33}
\definecolor{bred}{rgb}{.54,0,0}

\def\bred{\textcolor{\bred}}

\begin{document}
\renewcommand{\baselinestretch}{1.0}\normalsize
\begin{center}

{\large \bf What's in a ball?  Constructing and characterizing uncertainty sets

}
\renewcommand{\thefootnote}{\fnsymbol{footnote}}

\vspace{1cm}

\renewcommand{\baselinestretch}{1.0}

Thomas Kruse \footnote[2]{Department of Mathematics,
University of Duisburg-Essen,
Thea-Leymann-Str. 9,
D-45127 Essen, Germany;
Email: thomas.kruse@uni-due.de. \vspace*{0.3cm}}
        \hspace{1.5cm}
        Judith C. Schneider\footnote[1]{Finance Center Muenster,
       University of Muenster,
       Universit{\"a}tsstr. 14-16,
       D-48143 M{\"u}nster, Germany;
       Email: judith.schneider@wiwi.uni-muenster.de. \vspace*{0.3cm}}
 \hspace{1.5cm}
      Nikolaus Schweizer\footnote[3]
      {Department of Insurance and Risk Management,
       University of Duisburg-Essen,
        Lotharstr. 65, D-47451 Duisburg, Germany;
         Email: nikolaus.schweizer@.uni-due.de
}

\begin{quote}
\vspace*{1cm}
In the presence of model risk, it is well-established to replace classical expected values  by worst-case expectations over all
models within a fixed radius from a given reference model. This is the ``robustness'' approach. We show that previous methods for
measuring this radius, e.g. relative entropy or polynomial divergences, are inadequate for reference models which are moderately
heavy-tailed such as lognormal models. Worst cases are either infinitely pessimistic, or they rule out the possibility of fat-tailed
``power law'' models as plausible alternatives. We introduce a new family of divergence measures which captures intermediate
levels of pessimism.

\vspace*{0.5cm}


\vspace*{0.5cm}

\end{quote}

\end{center}
\

\renewcommand{\baselinestretch}{1.5}\small\normalsize

\section{Introduction}
When complex decisions are based on quantitative models, model uncertainty arises almost inevitably:
The models typically determine a set of probabilities for the events of interest,
and thus provide a quantification of \textit{risk}. Yet the model will never be perfectly accurate.
Therefore, decision-making has to  face the fact that some \textit{uncertainty} over probabilities remains.
Since the 1980s a series of contributions in the
operations research and control literature, see e.g. \citet{whittle1990risk,bertsimas2004price,ben2009robust},
suggest to incorporate the uncertainty into complex decision or optimization problems directly, i.e.,
to robustify the problem.
The basic idea of this approach is to augment calculations under a given reference model
by worst-case estimates taken over a collection of alternative models, the so-called uncertainty set. Typically,
the uncertainty set consists of all models which lie within a fixed radius from the reference model, i.e.,
in a given ball around the reference model.
To economists,
these ideas are best known through the work on robustness by Hansen, Sargent and coauthors in macroeconomics,
see \citet{hansen2011robustness} or through the ambiguity literature, e.g.  \citet{gilboa1989maxmin}.

The paper at hand attempts a careful reassessment of what is, arguably, the centerpiece of the robustness approach:
the choice of the divergence measure. In the vast majority of financial applications,
models correspond to continuous probability densities and thus to infinite dimensional objects.
Consequently, the choice of the divergence measure
is non-trivial, since the balls defined with respect to different divergences are non-equivalent.
A model which is at a finite distance from the
reference model for some divergence measure is often considered infinitely different under another.
In particular, the models taken into account as potential worst cases are different.

Our central question is how different choices of the divergence measure affect the degree of
uncertainty about the model's tail behavior included in the worst-case analysis.
In many applications, e.g. in finance and insurance, a misspecification of the model's tail behavior is viewed as
the most threatening form of model risk: Underestimating the heaviness of the tails leads to over-optimism about the distribution of
large losses. Moreover, especially when modeling heavy-tailed phenomena, it is very hard to estimate tail behavior
correctly from a limited amount of data \citep{clauset2009power}. Thus, model misspecification in the tails is very hard to rule out as a possibility.
We show that the classical divergence measures, Kullback-Leibler divergence (KL-divergence) and polynomial divergence, differ
widely in the way they allow for tail uncertainty.\par
Our first main contribution is to
provide a detailed assessment of these differences. The results can be summarized as follows:
Uncertainty sets defined in terms of KL-divergence allow for a broad range of different tail behaviors
when the model is light-tailed, e.g., when it has Gaussian or exponential tails. In particular, unless the reference model is
very light-tailed, uncertainty sets contain heavy-tailed alternative models. For models with heavier than exponential
tails such as power laws, (heavy-tailed) Weibull or lognormal distributions, the range of models taken into account as potential
worst cases is, in a sense, too wide. Even arbitrarily small KL-divergence balls contain alternative models without a finite first moment,
rendering the worst case expected value infinite. Unless one concedes that the expected value of the true data-generating process may indeed be infinite,
the worst case analysis is thus overly pessimistic.
Uncertainty sets defined in terms of polynomial divergence are well-suited for reference models which have a power law tail behavior
(e.g. Pareto distributions). For these models, the resulting uncertainty sets contain some power laws with heavier tails. Polynomial divergence balls
do not contain alternative models with qualitatively heavier tails. For instance, when the reference model is normal (or lognormal), the most heavy-tailed distributions
in the uncertainty set are normal (resp. lognormal) distributions with a slightly larger variance parameter.
Comparing the results for KL-divergence and polynomial divergence shows that neither choice of divergence is suitable for
distributions with heavier than exponential but lighter than power law tails. In applications, the most prominent examples of this class are the lognormal distribution and
(heavy-tailed) Weilbull distributions. \citet{watson2014approximate} and \citet{SS14} have proposed to solve this dilemma by truncating the
reference model. Yet the level of truncation always remains somewhat arbitrary and the truncated problem is not easy to handle from a numerical
point of view (see, again, \citet{SS14}). \par\par

Our second main contribution is an alternative solution to this problem. We propose a new family of divergences which
allow to conduct a worst case analysis for nominal models like Weibull or lognormal distributions, resulting in uncertainty regions which contain qualitatively more heavy-tailed models.. The new divergences lie between $\alpha$- and KL-divergence in the sense that they are finite more often than $\alpha$-divergence
but not quite as often as KL-divergence. Conversely, the new divergences lead to finite worst cases more often than KL-divergence and less often then $\alpha$-divergence.  The new divergences belong to the class of $F$-divergences  \citep{Csis63}. This implies that a rich body of
established theory is applicable to them.\footnote{See the survey by \citet{liese2006divergences} and the literature therein. } For instance, we rely on deep results for general $F$-divergences by
\citet{breuer2013measuring} to derive semi-explicit expressions for the worst-case distributions associated with
lognormal and Weibull distributions.

The paper is organized as follows: Section \ref{notation} introduces the key problem, the divergence measures
and the classes of distributions.
Section \ref{Ball} characterizes the contents of the uncertainty sets implied by the classical divergences: KL-divergence and polynomial divergence.
We proceed by presenting a new method to design tailor-made divergence measures for distributions which cannot suitably handled
with the classical divergences, see Section \ref{NewB}. Section \ref{Appl} derives the worst case distributions implied by our new divergences and gives a concrete numerical illustration in the context of space craft security. Section \ref{Con} concludes. All proofs are relegated to the Appendix.

\section{Concepts and Definitions}\label{notation}

In this section, we first introduce the problem of calculating expectations under model risk.
Here, we follow the classical robustness approach
as in, e.g., \citet{ben2013robust} or \citet{hansen2011robustness}. 
Afterwards, we introduce the basic
building blocks of our further analysis: A class of divergence measures which quantify how one model differs
from another, and three classes of reference models which cover most of the parametric families of distributions found in applications.

\subsection{The Problem}

There is a non-negative random variable $X$ and a decision-maker who works with the
assumption that $X$ is distributed according to some distribution $\nu$ which has continuous density  $f$.
The decision-maker is concerned about large values of $X$, i.e., $X$ could be thought of as the losses from some economic
endeavor. Throughout, our focus is on the expected value of $X$. To extend the scope, e.g., to expected disutility
from losses
or to expected losses above some threshold, one can simply redefine $X$ accordingly.

This distributional assumption on $X$ is called the \textit{reference model} or \textit{nominal model}.
Under the reference model, the mean of $X$ is thus given by
\[
E_{\nu}[X] = \int_0^\infty x f(x)dx.
\]
The decision-maker is concerned about the accuracy of his model and thus wishes to calculate the worst-case expected value of
$X$ over all alternative models $\eta$ with density $g$ which lie within a certain radius from the reference model,
\begin{equation}\label{wc}
\sup_{\eta: D(\eta | \nu)\leq \kappa } E_{\eta}[X] = \int_0^\infty x g(x)dx,
\end{equation}
where the radius $\kappa$ is a positive number and $D$ is a measure of the divergence of $\eta$ from $\nu$.

\subsection{Divergence Measures}

The choice of the divergence measure $D$ determines which alternative models are taken into account as possible alternative models.
Therefore, the validity of the worst case analysis strongly depends on a convincing choice of $D$.
The most popular
divergence in the robustness literature is Kullback-Leibler (KL) divergence, also known as relative entropy
\begin{align}
D^{KL}(\eta| \nu )=\int_0^\infty \log\left( \frac{g(x)}{f(x)}\right) g(x) dx,
\end{align}
due to its attractive statistical and decision-theoretic interpretations \citep{hansen2011robustness, watson2014approximate}.
When dealing with heavy-tailed nominal models,
KL-divergence has a a major drawback: It is too liberal as it includes implausible alternative models
with infinite expected values even in arbitrarily small divergence balls.
This problem was hinted at already in \citet{glasserman2012robust}, \citet{watson2014approximate} and \citet{SS14}.
We discuss it in detail in the next section.
Hence, one is forced to think about other divergence measures as an alternative to KL-divergence. As pointed out in the literature,
such alternatives may even reflect the decision-maker's preferences better \citep{friedman2007utility} or allow for an additional flexibility in the
worst-case analysis \citep{breuer2013measuring}.

In the following, we focus on divergence measures from the class of $F$-divergences \citep{Csis63}.
$F$-divergences are known to retain many of the theoretical properties of KL-divergence, see the survey by \citet{liese2006divergences}.
Moreover, a general method for calculating worst-case expectations for $F$-divergences is provided in \citet{breuer2013measuring}.\footnote{
	\citet{breuer2013measuring} actually cover the even broader class of Bregman divergences. Some authors, e.g., \citet{pflug2014multistage} have proposed a worst-case analysis
	based on entirely different types of probability distances such as the Wasserstein distance. Roughly speaking,
	when switching to Wasserstein distance one loses some of the statistical justifications of the worst-case approach but
	obtains a distance measure which is still applicable for very crude nominal models, such as modeling a continuous phenomenon by a model which is concentrated on finitely many points.}
Let $F:\mathbb{R^+}\rightarrow \mathbb{R}$ be a strictly convex function
with $F(1)=0$. Then the $F$-divergence between $\eta$ and $\nu$
is defined by the  integral
\begin{align}
D_F(\eta| \nu )=\int_0^\infty F\left( \frac{g(x)}{f(x)}\right) f(x) dx
\end{align}
if $\eta$ is absolutely continuous with respect to $\nu$ and by $D_F(\eta| \nu )= \infty$ otherwise.

Choosing $F(y)=y \log(y)$
recovers KL-divergence as a special case. A second special case are polynomial divergences $D^\alpha$,
also known as $\alpha$-divergences.
\citet{glasserman2012robust} propose to work with $\alpha$-divergences when dealing with heavy-tailed nominal models. These are defined by the choice
\[
F(y)=\frac{y^\alpha-1}{\alpha(\alpha-1)}
\]
where $\alpha>0$, $\alpha \neq 1$. R\'enyi divergences and Tsallis divergences (see, e.g., \citet{poczos2011estimation}) are monotone transformations of $\alpha$-divergence
and thus equivalent for our purposes. When $D^\alpha(\eta|\nu)$ is finite, it is well-known that in the limits $\alpha \downarrow 1$ and
$\alpha \uparrow 1$, we have $D^\alpha(\eta|\nu) \rightarrow D^{KL}(\eta|\nu)$. KL-divergence is thus, in a sense the limiting case $\alpha =1$
of polynomial divergence.

We make some more technical assumptions on the function $F$: Let $F:\mathbb{R^+}\rightarrow \mathbb{R}$ be continuously differentiable,
let $F(0) < \infty,$ and let $\lim_{y\rightarrow \infty} F(y)/y =\infty$. Assuming smoothness and finiteness
allows for a simple formulation of the worst case without the (direct) use of tools from convex analysis.
The final growth condition on $F$ is automatically satisfied for KL-divergence where $F(y)/y=\log(y)$.
As pointed above, KL-divergence is not restrictive enough when dealing with heavy-tailed models. We are therefore interested in
divergences which punish regions which are more likely under $\eta$ than under $\nu$ at least as much as KL-divergence.
Consequently, such divergences meet the growth condition as well. In particular,
we concentrate on $\alpha$-divergences with $\alpha>1$ in the following to satisfy this growth condition.

\subsection{Classes of Distributions}
We next introduce the three classes (i), (ii) and (iii) of models  considered in the remainder of the paper.
Our classification of models is not intended to be comprehensive, i.e., it is easy to construct models which fall in neither class.
Instead, we aim at tractable model classes which contain all the common parametric distributions on $\mathbb{R}_+$. We focus on distributions with unbounded support, in particular, we assume that there exists a threshold $x_f>0$ such that $f(x)>0$ for $x>x_f$, and analogously for $g$.

Throughout, it is useful
to write the densities $f$ and $g$ in exponential form (which is always possible)
\[
f(x)=\exp(- \varphi(x))\;\; \text{and}\;\; g(x)=\exp(- \gamma(x)).
\]
Roughly, our classification of models depends on growth conditions for the log-densities $\varphi$ and $\gamma$: Class (i) distributions correspond to linear or faster growth,
class (iii) distributions to logarithmic growth. Class (ii) distributions are those where the growth of the log-density lies between logarithmic and linear.
Qualitatively, the three classes correspond to (i) light-tailed models, (ii) models which are heavy-tailed but not fat-tailed, and
(iii) fat-tailed models.\footnote{There are many slightly conflicting
	notions of heavy-tailedness and fat-tailedness in the literature. A pair of (loose) definitions which is in line with our usage is to say
	that a distribution is heavy-tailed when it does not possess a finite moment-generating function, or when its tail decays slower than exponentially, and that it is fat-tailed
	when its tail behaves like a power law.}  In the following, we only spell out the definitions for $\nu$, the analogous ones apply to $\eta$.

We say that $\nu$ is a \textit{class (i) distribution} if there exists $c >0$ such that
$$\lim_{x \rightarrow \infty} \frac{\varphi(x) }{x} > c.$$ This class contains light-tailed distributions such as (one-sided) Gaussian distributions ($\varphi$ quadratic) and, as a boundary case,
exponential distributions ($\varphi$ linear).

We say that $\nu$ is a \textit{class (ii) distribution} if
$$\lim_{x \rightarrow \infty} \frac{\varphi(x) }{x} =0 \;\; \text{and}  \;\; \lim_{x \rightarrow \infty} \frac{\varphi(x) }{\log(x)} =\infty$$
and if there exists a function $\Phi:\mathbb{R}_+ \rightarrow \mathbb{R}$ with the following properties: $\Phi$ is positive, strictly concave, twice differentiable, and increasing over $[\bar{x}, \infty)$ for some $\bar{x}>0.$ Moreover, 
\[
\liminf_{x\to \infty}\frac {\Phi^{-1}(\varphi(x))}{x}>0 \quad \text{ and } \quad \limsup_{x\to \infty}\frac {\Phi^{-1}(\varphi(x))}{x}<\infty.
\]
A sufficient condition for the latter two properties which is satisfied in many examples is $\lim_{x\to \infty}\frac {\Phi^{-1}(\varphi(x))}{x}=1$.

This class contains distributions which are heavy-tailed but not fat-tailed.
Arguably, the most prominent examples are (heavy-tailed) Weibull distributions
for which $\varphi$ behaves like in $x^k$, $k \in (0,1)$ and log-normal distributions
for which $\varphi$ behaves like a quadratic polynomial in $\log(x)$ .

The function $\Phi$ should be thought of as a replacement of $\varphi$
which is monotonic, analytically simpler than $\varphi$  and less dependent on the distribution. $\Phi$ is an important building block in the construction of
our new divergences below. For illustration -- and as a preparation for later examples -- we spell this out for the Weibull and (generalized) lognormal case in detail.
In both cases, $\Phi$ is simply the leading term in the exponent of the density. The density  $f$ of a Weibull distribution with scale and shape parameter $\lambda>0$ and $k\in(0,1)$  is given by
\[
f(x)=\frac k\lambda \left(\frac x\lambda \right)^{k-1}e^{-\left(\frac x\lambda \right)^k}
\]
for $x\ge 0$ and $f(x)=0$ for $x\le 0$.\footnote{We assume $k\in (0,1)$ since Weibull distribution with $k\ge1$ are of class (i) rather than (ii). Similarly,
	we exclude generalized lognormal distributions with $r=1$ which belong to class (iii) distribution below.} In the Weibull case, we choose $\Phi(x)=(x/\lambda)^k.$
The density  $f$  of a generalized lognormal distribution is given by
\[
f(x)=\frac 1{ Z\cdot x }\exp\left(-\frac 1{r\sigma^r}|\log(x)-\mu|^r\right)
\]
with $Z= 2 r^{1/r}\sigma \Gamma(1+1/r)$, $r > 1$, $\sigma>0$ and $\mu \in \mathbb R$.
In this case, we choose $\Phi(x)=\frac 1{r\sigma^r}\log(x)^r$. The generalized lognormal distribution with $r=2$ is the usual lognormal distribution.
For further discussion and applications of the generalized case, see \citet{kleiber2003statistical}.

Finally, we say that $\nu$ is a \textit{class (iii) distribution} if there exists $c < \infty$ such that
\begin{equation}\label{defiii}
\lim_{x \rightarrow \infty } \frac{\varphi(x)}{\log(x)} =c.
\end{equation}
This class contains only distributions with fat, polynomial tails such as Pareto distributions.
In particular, we say that a distribution
has polynomial tails of degree $c$ if \eqref{defiii} holds for this particular value of $c$.

\section{What's in a ball: Classical Divergence Measures}\label{Ball}

In this section, we analyze how the different divergence measures influence the worst-case analysis, i.e., which alternative
distributions are taken into account as potential worst-case models. In particular, we are interested in the ability of divergence measures
to capture model uncertainty concerning the tail behavior of the nominal model. There are at least two motivations for this focus:
First, model misspecification in the tail is arguably the major concern when thinking about model risk, e.g., in financial
applications. For instance, the lognormal assumption behind the Black-Scholes model is frequently criticized even in the popular press
since it might underestimate the heaviness of tails.
Second, unless the data generating process is rather light-tailed (e.g. Gaussian),
it is very hard to accurately estimate the tail behavior from a limited amount of data.
Consequently, when one heavy-tailed model like a lognormal, Weibull
or power law model is chosen, one typically  cannot confidently rule out other types of heavy-tailed models as
plausible alternatives, see \citet{clauset2009power} for a discussion and empirical examples.
Therefore, we focus on the question how well different divergence balls can handle uncertainty about the type of tail behavior.

More specifically, we are interested in how rich the
content of a ball with fixed radius is for different combinations of nominal model and divergence measure.
The next result shows that this question is essentially equivalent to studying the following simpler question:
How rich is the set of models at arbitrary finite distance from the nominal model?
For any distribution $\eta$ within a finite $F$-divergence from $\nu$ there exists a distribution $\tilde{\eta}$
which has the same tail behavior as $\eta$ and a distance smaller than $\kappa$ from $\nu$.
Thus, with respect to tail behavior,
balls with a small radius are as rich as balls with a large radius. This result is summarized in the following proposition:

\begin{proposition}\label{prop small balls}
	Fix $\kappa>0$ and $\nu$ and $\eta$ with $D_F(\eta|\nu)< \infty$ for some $F$ which satisfies our standing assumptions.
	Suppose that $f(x)$ is positive for $x$ above some threshold and that $\eta$ has heavier tails than $\nu$ in the sense that
	$\lim_{x \rightarrow \infty} f(x)/g(x) =0$. Then there exists a distribution $\tilde{\eta}$ with density
	$\tilde{g}(x)=\exp(-\tilde{\gamma}(x))$ with $D_F(\tilde{\eta}|\nu) < \kappa$ and with the same tail behavior as $\eta$,
	\[
	\lim_{x \rightarrow \infty} \frac{\tilde{\gamma}(x)}{ \gamma(x)}=1.
	\]
\end{proposition}

In the following, we first consider KL-divergence, the most prominent
divergence measure in the literature. 
We then proceed to polynomial divergences, the alternative for heavy-tailed models
proposed by
\citet{glasserman2012robust} and characterize in detail their scope and limitations.

The key results of the remainder of this section can be summarized as follows:
$KL$-divergence is suitable for class (i) distributions and $\alpha$-divergence
(with a well-chosen value of $\alpha$)
is suitable for class (iii) distributions. For class (ii) distributions neither of the two works
since, roughly speaking, KL-divergence is too liberal and $\alpha$-divergence is too restrictive.

\subsection{KL-Divergence Balls}

The next proposition characterizes the contents of KL-divergences balls and shows that they are fairly rich.

\begin{proposition}\label{propKL}
	Let $\nu$ and $\eta$ be two probability distributions with associated functions $f$, $\varphi$, $g$ and $\gamma$. Let $\int_0^\infty \gamma(x) g(x) dx < \infty$. Then $D_{KL}(\eta|\nu)$
	is finite if and only if
	\[
	\int_0^\infty \varphi(x) g(x) dx < \infty.
	\]
\end{proposition}

The condition $\int_0^\infty \gamma(x) g(x) dx < \infty$ states that the integral of $\gamma$ with respect to $\eta$ is finite.
This is a
mild regularity condition on the alternative distribution $\eta$, equivalent to the postulate that $\eta$ has finite entropy.
The condition $\int_0^\infty \varphi(x) g(x) dx < \infty$ links the nominal model to the worst-case model.
It shows that KL-divergence balls are well-suited for a worst-case analysis around class (i) nominal distributions,
and not well-suited for classes (ii) and (iii).
For the boundary case, i.e., $\nu$ being an exponential distribution, the condition is essentially equivalent to the following: $\eta$
has a finite KL-divergence from $\nu$ whenever $\eta$ has a finite expected value. For other class (i) distributions, existence
of a finite expected value is replaced by stronger moment conditions. For instance, if $\nu$ is Gaussian then $\varphi$ is quadratic, implying that $\eta$
can only have finite KL-divergence from $\nu$ if $\eta$ possesses a second moment. Consequently, KL-divergence balls around class (i) models can
contain possible worst case distributions with fairly heavy tails. Unless $\nu$ is \textit{very} light-tailed, these balls still contain
a rich collection of distributions with polynomial tails.

For class (ii) and (iii) distributions, the condition is weaker than any moment condition, postulating integrability of a function, $\varphi$,
which grows less than linearly with respect to $\eta$: the condition is weaker than $E_{\eta}[X]<\infty$.
Combining this observation with Proposition \ref{prop small balls} implies that
even small KL-divergence balls around class (ii) and (iii) distributions contain models which do not possess an expected value.
This renders the worst case problem \eqref{wc} trivial, the worst case being infinitely bad.

\subsection{$\alpha$-Divergence Balls}

We next characterize the contents of $\alpha$-divergence balls for the case $\alpha>1$.
The main finding of the next proposition is that
$\alpha$-divergence can only be finite if the ratio of $\gamma$ and $\varphi$
is bounded from below, implying that $\gamma$ must not grow much slower than $\varphi$.
For technical reasons, the proposition differentiates between class (iii) models and more light-tailed models
but the main finding is the same in both cases.

\begin{proposition}\label{proppoly}
	\item[(i)] Let $\nu$ be a class (i) or (ii) distribution with associated functions $f$ and $\varphi$ where $f$ is bounded. Let $\eta$ be another distribution with associated
	functions $g$ and $\gamma$. Suppose that the limit $\lim_{x \rightarrow \infty} \frac{\varphi(x)}{\gamma(x)}$ exists and that $g(x)/f(x)$ is bounded on any compact interval.
	Then
	\[
	\lim_{x \rightarrow \infty} \frac{\gamma(x)}{\varphi(x)} > \frac{\alpha-1}{\alpha}
	\]
	implies $D_{\alpha}(\eta|\nu)< \infty$, and
	\[
	\lim_{x \rightarrow \infty} \frac{\gamma(x)}{\varphi(x)} < \frac{\alpha-1}{\alpha}
	\]
	implies $D_{\alpha}(\eta|\nu)= \infty$.
	\item[(ii)] Let $\nu$ be a class (iii) distribution with associated functions $f$ and $\varphi$ where $f$ is bounded and $\lim_{x\rightarrow \infty} \varphi(x)/\log(x)=c>1$. Let $\eta$ be another distribution with associated
	functions $g$ and $\gamma$. Suppose that the limit $\lim_{x \rightarrow \infty} \frac{\varphi(x)}{\gamma(x)}$ exists and that $g(x)/f(x)$ is bounded on any compact interval.
	Then
	\[
	\lim_{x \rightarrow \infty} \frac{\gamma(x)}{\varphi(x)} > \frac{\alpha-1}{\alpha}+\frac{1}{c\, \alpha}
	\]
	implies $D_{\alpha}(\eta|\nu)< \infty$, and
	\[
	\lim_{x \rightarrow \infty} \frac{\gamma(x)}{\varphi(x)} < \frac{\alpha-1}{\alpha}+\frac{1}{c\, \alpha}
	\]
	implies $D_{\alpha}(\eta|\nu)= \infty$.
\end{proposition}

From the proposition, we see that $\alpha$-divergence is a superior alternative to KL-divergence if $\nu$ is a class (iii) model:
For class (iii) models, i.e., models with polynomial tails, $\varphi$ behaves essentially like $c \log(x)$ where $c$
is the parameter of polynomial decay. By the proposition, we can thus conclude that $\alpha$-divergence balls around $\nu$
contain models with any qualitatively lighter tail behavior than $\nu$, as well as models with heavier polynomial tails. Precisely, the ball contains
class (iii) models whose polynomial decay rate $\tilde c$ satisfies
\[
\tilde{c} > \frac{\alpha-1}{\alpha} c + \frac{1}{\alpha}.
\]
In particular, the parameter $\alpha$ can be used to control the amount of model uncertainty with regard to tail behavior.

For class (i) and (ii) models, considering polynomial divergence balls in a worst-case analysis is essentially the same as postulating that, qualitatively, the nominal model
does not underestimate tail risk: From part (i) of the proposition, we see that the $\alpha$-divergence balls around a Gaussian model can contain models with slightly heavier
Gaussian tails - but nothing qualitatively heavier such as exponential, log-normal, or polynomial tails. Similarly, an $\alpha$-divergence ball around a log-normal
model contains other models with, possibly slightly heavier, log-normal tails but no models with polynomial tails. We thus see
that $\alpha$-divergence cannot successfully capture uncertainty about the heaviness of the tails for class (i) and (ii) models.

\subsection{Discussion}

Finally, let us comment on the dependence on $\alpha$. Inspecting Proposition \ref{proppoly}, we notice
that the conditions for finiteness of $D^{\alpha}$ become more restrictive as $\alpha$ increases. As mentioned above,
the limit $ \alpha \downarrow 1$ corresponds, in a sense, to KL-divergence. The limit $\alpha \uparrow \infty$ is rather well-studied as well:
A straightforward calculation shows that $D^{\infty}(\eta|\nu)$ is an increasing function of the supremum of $g(x)/f(x)$.
In this case, balls of a fixed radius correspond to sets of models with $g(x)/f(x) \leq C$ for some $C$. Worst case densities
within such sets equal zero up to some threshold and equal
$C\times f(x)$ above the threshold. This shows that the worst case associated with $D^{\infty}(\eta|\nu)$ is the so-called Conditional Value at Risk (CVaR) associated with a
given quantile. Indeed, the literature on risk measures \citep{follmer2011stochastic} has highlighted that any coherent risk measure
possesses a robustification, i.e., it can be formulated as a worst-case expectation over \textit{some} convex set of alternative models.
For CVaR, arguably the most prominent coherent risk measure,
we have just seen that this convex set is a ball with respect to $D^{\infty}$.
Interpreting the associated worst-case distribution as an alternative model nicely illustrates the restrictiveness of
$\alpha$-divergence balls: The worst-case distribution may look quite different, i.e., it may be zero over substantial parts of the
support of the nominal model. Yet it sticks closely to the nominal model in the sense that it always inherits its tail behavior.

The fact that there is a marked qualitative difference between $\alpha$-divergence and KL-divergence may seem puzzling
since $D^{\alpha}(\eta|\nu)$ converges to $D^{KL}(\eta|\nu)$ as $\alpha \downarrow 1$.
To understand this apparent contradiction, notice that this statement about convergence only makes sense when  $D^{\alpha}(\eta|\nu)$  is finite
for $\alpha>1$. The difference between KL- and $\alpha$-balls stems from distributions which have finite KL-divergence but infinite $\alpha$-divergence.
In the next section, we propose new divergences  which lie between $\alpha$- and KL-divergence in the sense that they are finite more often than $\alpha$-divergence
but not quite as often as KL-divergence.

In sum, for worst-case analysis of a nominal model from class (i) we can confidently use KL-divergence, for class (iii) models
we can use $\alpha$-divergence with a suitable value of $\alpha$. However, for class (ii) models neither of these approaches is satisfactory with KL-divergence overstating model uncertainty and $\alpha$-divergence
understating it. This is the problem we address in the remainder of the paper.

\section{What's in a ball: Construction of new divergence measures for worst case analysis}\label{NewB}

We have seen that neither polynomial divergences nor KL-divergence are particularly suitable for constructing uncertainty regions around class (ii) models. In this section we propose a generic construction method for alternative $F$-divergences which are tailored to a given class (ii) model. The resulting uncertainty regions are rich enough to contain some more heavy-tailed models, including some power laws, but they are still sufficiently restrictive to exclude models without finite mean from the worst case analysis. More specifically,  these divergences are thus, e.g., well-suited for a worst-case analysis of log-normal risks which takes
into account the possibility that the true tail behavior is polynomial. Likewise, they allow to construct a ball around a Weibull model which includes models with lognormal tails -- without implying an infinite worst case.

Recall that in the definition of class (ii) distributions we assumed existence of a function $\Phi$ with the same asymptotic behavior as $\varphi$.
Moreover, we assumed that $\Phi$ is positive, twice differentiable, strictly concave, and strictly increasing over the interval $[\bar{x}, \infty)$.
We denote by $\Phi^{-1}:[\Phi(\bar{x}),\infty)\rightarrow \mathbb{R}$ the inverse of the restriction of $\Phi$ to $[\bar{x}, \infty)$
and observe that $\Phi^{-1}$ is positive, twice differentiable, strictly convex, and strictly increasing.

We begin with a heuristic derivation of the class of divergences we introduce. For the moment, let us fix some nominal model $\nu$ from class (ii). Suppose we want to construct an $F$-divergence such that $F$-divergence balls around $\nu$ contain, roughly,
those alternative models which possess a finite moment of order $\theta$ but not necessarily higher moments. We thus try to achieve an  equivalence between
\[
D_F(\eta | \nu)= \int_0^\infty F\left(\frac{g(x)}{f(x)} \right) f(x) dx < \infty \;\;\text{ and }\;\;  \int_0^\infty x^\theta g(x) dx<\infty.
\]
We claim that a good candidate for $F$ is $F( y)=y \Phi^{-1}(\log(y))^\theta$ so that
\[
D_F(\eta | \nu)= \int_0^\infty  \Phi^{-1}\left(\log\left( \frac{g(x)}{f(x)}   \right)\right)^\theta g(x) dx.
\]
This claim is based on the following reasoning: If $g$ does not possess all moments, it must decrease much slower than $f$. Thus, the tail behavior of $g/f$ in the argument of $F$ is essentially the same as the behavior of $1/f$. This suggests the following equivalence in tail behavior
\[
\Phi^{-1}\left(\log\left( \frac{g(x)}{f(x)}   \right)\right)^\theta \sim \Phi^{-1}\left(\log\left( \frac{1}{f(x)}   \right)\right)^\theta \sim
\Phi^{-1}(\phi(x))^\theta \sim x^\theta,
\]
which implies that the integrand in $D_F(\eta | \nu)$ should behave like $x^\theta g(x)$ for large $x$.

In the case $\Phi(x)=x$ and $\theta=1$, the above choice of $F$ simply recovers KL-divergence. For $\Phi(x)=c \log(x)$, the resulting divergence is a polynomial divergence. The nominal models leading to these two choices of $F$ are an exponential distribution (class (i)) and a power law (class (iii)). For class (ii) nominal models, the resulting $\Phi$ and $F$ lie somewhere between these cases regarding their tail behavior. For instance, in the lognormal case with $\sigma^2=1/2$ we have $\Phi(x)= \log(x)^2$ which implies
\[
F(y)= y \exp\left(\theta \sqrt{\log(y)}\right).
\]
Indeed, this function displays a growth behavior between the $y \log(y)$ of KL-divergence and the $y^{\alpha}$, $\alpha >1$, of polynomial divergence. Yet the function is only well-defined for $y \geq 1$. To circumvent problems of this type, the alternative divergences we propose are defined piece-wise. For large $y$ we follow the above construction. For small $y$, i.e., in regions which are not considered \textit{much} more likely by $g$ than by $f$, we choose $F$ as in a KL-divergence.

\begin{definition}\label{divergences}
	Let $\nu$ be a class (ii) distribution with associated functions $f$, $\varphi$, $\Phi$.
	Let $\theta > 1$ and $\bar{y}= \exp(\Phi(\bar{x}))$.
	The function $F_{\Phi}$ is defined by
	\[
	F_{\Phi}(y)=\left\{
	\begin{array}{ll}
	y \log(y) & \text{ for } y \leq \bar{y}\\
	y a \Phi^{-1}(\log(y))^\theta+b & \text{ for } y > \bar{y}
	\end{array}
	\right.
	\]
	with
	\[
	a=\frac{1+\log(\bar{y})}{\Phi^{-1}(\log(\bar{y}))^\theta+ \theta \Phi^{-1}(\log(\bar{y}))^{\theta-1} {\Phi^{-1}}'(\log(\bar{y}))}
	\]
	and
	\[
	b= \bar{y} \log(\bar{y})- a \bar{y} \Phi^{-1}(\log(\bar{y}))^\theta
	\]
	
\end{definition}
The next lemma verifies that $F_{\Phi}$ fulfills the assumptions made when introducing $F$-divergences. The $F$-divergence $D_{F_\Phi}$ based on $F_{\Phi}$ is thus well-defined.
\begin{lemma}\label{lem42}
	$F_{\Phi}$ is strictly convex, continuously differentiable and satisfies $F_{\Phi}(1)=0$ and $\lim_{y\rightarrow \infty} F_{\Phi}(y)/y =\infty.$
\end{lemma}

For class (ii) nominal models, we next show that finiteness of some $\alpha$-divergence implies finiteness of $F_\Phi$-divergence, which, in turn, implies finiteness of KL-divergence. In light of Proposition \ref{prop small balls}, this shows that, as intended,
$F_\Phi$-divergence balls are richer than $\alpha$-divergence balls but not as rich as KL-divergence balls.

\begin{proposition}\label{ordering}
	Let $\nu,\eta$ be two distributions and assume that $\nu$ is of class (ii). Then $D_{F_\Phi}(\eta|\nu)<\infty$ implies $D^{KL}(\eta|\nu)<\infty$. Moreover, we have $D_{F_\Phi}(\eta|\nu)<\infty$, whenever $D^\alpha(\eta|\nu)<\infty$ for some $\alpha>1$.
\end{proposition}

The next two propositions characterize which alternative models are included in $F_\Phi$-divergence balls -- and which are not. Under some weak regularity conditions on the density $g$, we find that $F_\Phi$-divergence indeed manages to include those models in the balls which have a finite $\theta$-th moment, and to exclude those which do not possess this moment. We begin by showing that $F_\Phi$-divergence balls are indeed as rich as intended:

\begin{proposition}\label{suff_cond_finite_div}
	Let $\nu$ be a class (ii) distribution with associated functions $f$, $\varphi$, $\Phi$.
	Let $\theta \geq 1$ and let $\eta$ be a distribution with density $g$ such that $g(x)\le 1$ for all $x$ large enough and such that $g/f$ is bounded on any compact interval. If $E_{\eta}[X^\theta]< \infty$ then $D_{F_\Phi}(\eta|\nu)< \infty$.
\end{proposition}

Proposition \ref{suff_cond_finite_div} shows that $F_\Phi$-divergence balls include alternative models with a polynomial decay rate above $\theta$. We now turn to the converse direction and verify that all sufficiently regular models whose tails are heavier than a power-law with some infinite moment of order $t<\theta$ are excluded from any divergence ball around the nominal model.

\begin{proposition} \label{prop44}
	Let $\nu$ be a class (ii) distribution with associated functions $f$, $\varphi$, $\Phi$ and assume one of the following
	\begin{itemize}
		\item[\textnormal{(i)}] $\Phi^{-1}$ is pseudo-regularly varying\footnote{\label{def_prv}Following \cite{buldygin2002properties} a measurable function $h\colon \mathbb R_+\to (0,\infty)$ is called pseudo-regularly varying if
			\[
			\limsup_{c\to 1}\limsup_{x\to \infty}\frac {h(cx)}{h(x)}=1
			\]
		},
		\item[\textnormal{(ii)}] $\nu$ is a Weibull distribution with $k \in (0,1)$,
		\item[\textnormal{(iii)}] $\nu$ is a generalized lognormal distribution with $r >1$.
	\end{itemize}
	Let $\theta > 1$ and let $\eta$ be a distribution with density $g$ and assume that $\liminf_{x\to \infty} x^{t+1} g(x)>0$ for some $t \in (1, \theta)$. Then we have $D_{F_\Phi}(\eta|\nu)=\infty$.
\end{proposition}

In the proposition, the Weibull case (ii) is a direct consequence of case (i). The lognormal case (iii) follows from a weaker sufficient condition given in the proof. The proposition is not strong enough to imply that a ${F_\Phi}$-divergence ball cannot contain models without a finite $\theta$-th moment: A priori, there could be such models which violate the regulatory conditions of proposition \ref{prop44}, e.g., through an oscillatory behavior in the tail. In the  next section we  prove that the worst case expected values taken over  ${F_\Phi}$-divergence balls are indeed finite.

Clearly, the thus defined divergences depend on $\nu$ through the function $\Phi$.
On the one hand, this implies, that for all distributions with the same function $\Phi$ we can apply the same divergence.
On the other hand, it is necessary to tailor the divergence to the considered nominal model to some extent.
We believe that this is indeed unavoidable in the subexponential case -- a similar issue arises in the form of choosing the parameter
$\alpha$ for $\alpha$-divergence as well.

In the case of $\alpha$-divergence, the choice of $\alpha$ actually serves a double role, adapting the divergence to the nominal model
\textit{and} deciding on the heaviest tail behavior to be included (within the limitations of $\alpha$-divergence discussed above):
$\alpha$ essentially determines the maximum difference in tail behavior (measured in multiples of $\varphi$).
For our $F_\Phi$-divergences, the two choices are distinct: The function $F_\Phi$ is tailored to the distribution
while the choice of $\theta$ quantifies model risk in the sense of specifying the heaviest polynomial tail behavior that is taken into account in the worst-case analysis.

\section{Application to worst-case analysis}\label{Appl}

In this section, we derive a condition which ensures that the worst case problem for $F_\Phi$-divergence has a finite solution and provide an expression for the worst case. To achieve this, we build on results of \citet{breuer2013measuring}. The one major difference to their more general approach is that our sufficient condition and worst case density can be formulated rather explicitly, in particular avoiding the machinery of convex analysis.

\begin{proposition}\label{prop51}
	Let $\nu$ be a class (ii) distribution with associated functions $f$, $\varphi$, $\Phi$ and fix $\theta>1$. Define $\psi(x)=a\Phi^{-1}(x)^\theta$.
	If for all $\alpha_1 \in \mathbb{R}$ and $\alpha_2 >0$ there exists $c>0$ such that
	\begin{equation}\label{condIc}
	I(c) = \int_{c}^\infty e^y(\psi(y)+\psi'(y))(\psi'(y)+\psi''(y))f\left(\frac{F_\Phi'(e^y)-\alpha_1}{\alpha_2}\right)dy%<\infty
	\end{equation}
	is well-defined and finite, then the worst case in \eqref{wc} is finite and there exists $(\alpha_1^{wc},\alpha_2^{wc})\in \mathbb{R}\times \mathbb{R_+}$ such that the worst case model $\eta^{wc}$ has the density  $$g^{wc}(x)=(F_\Phi')^{-1}(\alpha_1^{wc}+\alpha_2^{wc}x)f(x)$$ where $(F_\Phi')^{-1}$ denotes the inverse of the derivative of $F_\Phi$. Moreover, $(\alpha_1^{wc},\alpha_2^{wc})\in \mathbb{R}\times \mathbb{R_+}$ are uniquely characterized by the conditions that $g^{wc}$ integrates to $1$ and that $D_{F_\phi}(\eta^{wc}|\nu)=\kappa$.
\end{proposition}

The requirement that $I(c)$ is well-defined simply alludes to the fact that $y$ must be sufficiently large to ensure that $(F_\Phi'(e^y)-\alpha_1)/\alpha_2$ is positive and thus a valid argument for $f$. We next verify that Proposition \ref{prop51} is applicable in our two running examples, Weibull distributions and generalized lognormal distributions:

\begin{corollary} \label{cor52}
	Let $\nu$ be a class (ii) distribution with associated functions $f$, $\varphi$, $\Phi$ and assume one of the following
	\begin{itemize}
		\item[\textnormal{(i)}] $\nu$ is a Weibull distribution with $k \in (0,1)$,
		\item[\textnormal{(ii)}] $\nu$ is a generalized lognormal distribution with $r >1$.
	\end{itemize}
	Then, for all $\theta >1$,  $\alpha_1 \in \mathbb{R}$ and $\alpha_2 >0$ there exists $c>0$ such that $I(c)$ from \eqref{condIc} is well-defined and finite. In particular, the worst case in \eqref{wc} is finite and given by Proposition \ref{prop51}.
\end{corollary}

In Section \ref{numerik}, we demonstrate that Proposition \ref{prop51} is explicit enough to make calculating the worst cases a straightforward numerical task. However, the proposition does not provide an easy to interpret expression for the worst case. For Weibull and lognormal distributions, the next proposition gives a result of this type, providing simple closed-form expressions of functions which are asymptotically equivalent to the worst case density.

\begin{proposition}\label{prop53}
	Let $(\alpha_1,\alpha_2)\in \mathbb{R}\times \mathbb{R_+}$ and $g(x)=(F_\Phi')^{-1}(\alpha_1+\alpha_2x)f(x)$.
	\item[(i)] If $\nu$ is a Weibull distribution with $k \in (0,1)$ then  $g$ is asymptotically equivalent to $$\overline{g}(x)=f(x)\exp\left(%(-\left(\frac x\lambda\right)^k+
	\left(\frac {\alpha_1+\alpha_2x}{a\lambda}\right)^{\frac k \theta}\right).$$
	\item[(ii)] If $\nu$ is a lognormal distribution, i.e., a generalized lognormal with $r=2$, then $g$ is asymptotically equivalent to $$\overline{g}(x)=f(x)\exp\left(\frac 1{2\sigma^2 \theta^2}\log\left(\frac {\alpha_1+\alpha_2x}{a}\right)^2\right).$$
\end{proposition}

Thus, even though the uncertainty regions contain much more heavy-tailed models, the worst case densities display a tail behavior which is qualitatively similar to the respective nominal models. In the lognormal case, the exponent still behaves like a quadratic polynomial in $\log(x)$. Likewise, the exponent in the Weibull case is still a (fractional) polynomial with maximal degree $k$. This is reminiscent of the fact that, under KL-divergence, the worst case for a normal distribution remains a normal distribution, and the worst case for an exponential distribution remains an exponential distribution.

\subsection{Numerical Illustration}\label{numerik}
This section aims to give a numerical illustration of the newly constructed divergence. The new divergence is designed for nominal models of class (ii), as our running examples the lognormal and the Weibull distribution. The lognormal distribution is heavily applied for basic model description in finance or insurance and is the default for  standard models of loss description under Solvency II and in Basel III \citep{hurlimann2009non,frachot2004loss}.  The Weibull distribution is adopted not only in insurance but also in many applications in the natural sciences, see \citet{kleiber2003statistical} and \citet{clauset2009power}. The following example is inspired by the solar flare  or solar particle event  literature where worst case scenario analysis is a common method and where Weibull distributions are widely used reference models \citep{townsend2003carrington}. The example underlines once more the broad applicability of the divergence approach within various areas of economics or operations research. It also highlights  the need for a well-suited divergence measure for class (ii) models. \par

We choose the parameters of the Weibull distribution $(k,\lambda)=(0.4015,0.6821)$ from fits to a particular solar flare event, see \citet{xapsos2000characterizing} Table 1, column 2. A good grasp of how severe such an event can be under model risk is crucial for  planning space crafts and for assessing the safety of crews in space \citep{townsend2006carrington}. As we are mainly interested in the worst case severity of
extreme events, we choose as the reference model not the Weibull distribution itself, but the Weibull distribution conditioned on
realizations above its  95\% quantile $q_{0.95}=10.4878$.\footnote{Notice that the worst-case densities we derived for the Weibull case apply to this truncated Weibull distribution as well.  To determine the new divergence we have chosen $\bar y$ from Definition \ref{divergences} as $\bar y=\exp(1)$.}

\begin{figure}[t]
	\begin{center}
	\end{center}
	\includegraphics[width=\textwidth]{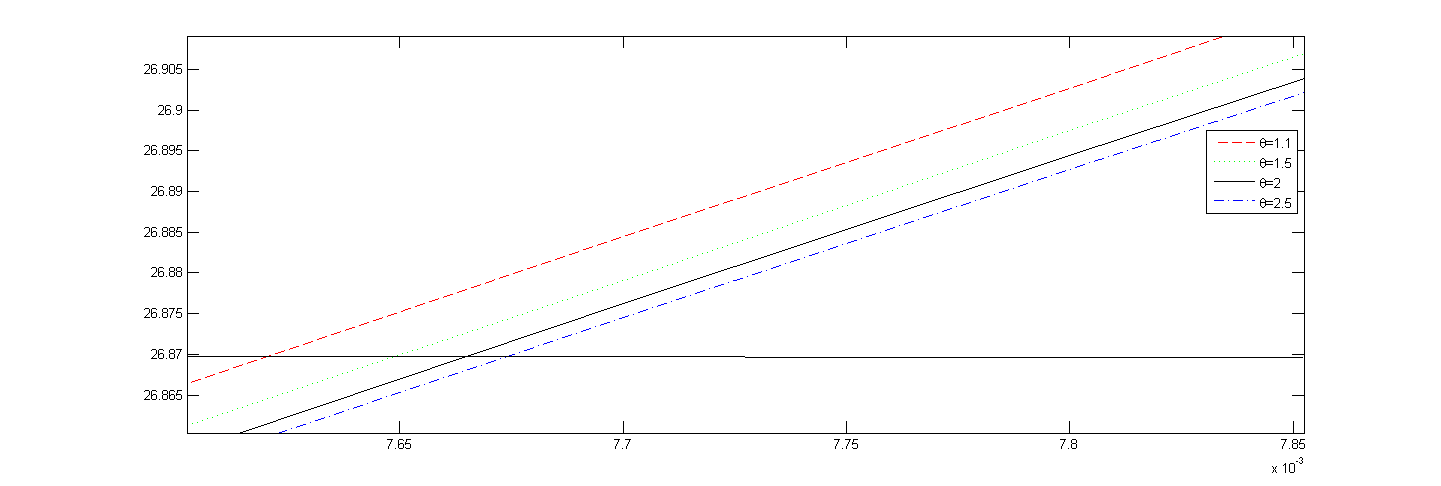}\vspace*{2mm}
	
	\vspace{1ex}
	\caption{\footnotesize{The figure illustrates the worst-case expected value for $\theta\in \{1.1, 1.5, 2.0, 2.5\} $  for a range of $\kappa$ values from $0.006$ to  $ 0.00788.$}}
	\label{fig:Theta}
\end{figure}
Figure \ref{fig:Theta} shows the worst case expected value (of the Weibull distribution conditioned above its 95\% quantile) as a function of the radius $\kappa$ of the divergence ball and the parameter $\theta$ of the new divergence. Recall that $\theta$ resembles a moment condition for the potential worst case models. For the same $\theta$ the worst case expected value is an increasing function in $\kappa.$ Vice versa, for the same $\kappa$ the worst case is an increasing function in $\theta.$\footnote{ This is due to the fact that in this example $F_{\Phi}(y)$ is increasing in $\theta$ for all $y$.} This implies that  different $(\theta,\kappa)$-tuples can lead to the same worst case value. This is indicated by the horizontal line in Figure \ref{fig:Theta}. Therefore, it is of great importance to find economically sensible choices for both variables. A risk engineer usually has an idea what a realistic $\theta$ is, i.e., whether a sensible alternative model posses a finite $\theta$-th moment or not. The question how to choose $\kappa$ is more intricate, see \citet{SS14}.  Herein we choose the following calibration: For our (truncated) Weibull nominal model, KL-divergence worst cases are infinite while $\alpha$-divergence worst cases still exist. However,  they are very restrictive concerning the tails of  potential worst-case models and thus underestimate the model risk the risk manager is exposed to. We take the $\alpha$-divergence worst cases as benchmarks and determine the radius $\kappa$ in the new divergence between the reference model and this benchmark. More specifically, we calculate $\alpha$-divergence worst cases for varying $\alpha$ such that the worst case  for every $\alpha$ is exactly $10\%$ above the nominal expected value. We call this the 10\% safety margin under $\alpha$-divergence. For varying $\theta$, the resulting $\kappa$'s are then used to calculate expected worst case values under the new divergence. Table 1 reports the difference between the $\alpha-$divergence safety  margin  (10\%) and the worst case safety margin of the new divergence.  We have seen in Figure \ref{fig:Theta} that for a given $\kappa$ the worst-case expected value is an  increasing function in $\theta$ and vice versa. In the table we observe that  for a given $\theta$ the safety margin is not monotonic in $\alpha.$ Recall that $\alpha$-divergence balls place rather strict conditions on the  tail behavior of the potential worst case models. The smaller $\alpha$, the more heavy-tailed the models can potentially be. We have, however, fixed the $\alpha$-worst case to result in a safety margin of  10\%  for every $\alpha$. This  in turn leads to different sizes of the balls.  Moreover, we also observe a non-monotonicity for varying $\theta$ given one particular $\alpha.$ This is caused by the relation between different  $(\theta,\kappa)$-tuples and the worst case expected value, see Figure \ref{fig:Theta}. While the latter is fixed if the risk manager has an idea about the existence of moments of the alternative distribution, the first is less clear. Although $\alpha$ in general allows a certain control of the model uncertainty with regard to tail behavior, its  strict conditions on the potential worst case distributions are rather opaque. Thus, how to set $\alpha$ is less clear and enters in our illustration due to the way we calibrated the size of the ball. For more discussion of alternative methods to determine the radius of the ball, see \citet{SS14} and \citet{watson2014approximate}.   But what can be stated is that the table nicely illustrates that the $\alpha$-divergence worst case can substantially underestimate model risk. Our alternative divergence measure fills the gap and gives a much more sensible idea of the amount of model risk within the application.

\renewcommand{\arraystretch}{1.5}
\begin{table}
	\begin{center}
		\begin{tabular}{|c|c|c|c|c|c|c|c|c|}
			\hline
			$ \theta\; \backslash \; \alpha $ & $    1.1$ & $    1.5$ & $      2$ & $    2.5$ & $      3$ \\ \hline
			$1.1$ & $ 1.0120$ & $ 0.1220$ & $0.1111$ & $0.2418$ & $0.4042$ \\ \hline
			$1.5$ & $3.0959$ & $0.3175$ & $0.1282$ & $0.2315$ & $0.3884$ \\ \hline
			$  2$ & $10.9746$ & $0.7724$ & $0.1772$ & $ 0.2310$ & $0.3792$ \\ \hline
			$2.5$ & $34.2643$ & $  1.6900$ & $0.2611$ & $0.2378$ & $0.3756$ \\ \hline
		\end{tabular}
	\end{center}\vspace*{0.5 cm}\caption{Safety margins in percent above the reference safety margin for different $\alpha$ and $\theta$. The reference safety margin under $\alpha$-divergence is $10\%$.
	Distribution conditioned on being above the $95\%$-quantile $q_{0.95}=10.4878$. The nominal mean is $24.1715$.
}
\end{table}\label{Tab:1}

\section{Conclusion}\label{Con}
This paper contributes to what may be called ``divergence'' approach to model uncertainty: Expected values under a reference model are augmented by worst-case expectations calculated over potential alternative models which lie within a uncertainty set. The radius of the uncertainty set is fixed and usually defined by a so-called divergence measure. The divergence measure is the centerpiece of this approach. A model which my be taken as an alternative model under one particular divergence measure, might not be considered under another divergence measure. Therefore, it is of great importance to understand how different choices of divergence measures affect the degree of uncertainty. We focus, in particular, on the tail behavior of the alternative and potential worst case distributions under two prominent divergence measures: KL-divergence and polynomial divergence. We highlight the marked qualitative difference between KL-divergence and  polynomial divergence for three different classes of nominal models. For class (ii) models -- heavier than exponential but lighter than power law tails -- polynomial divergence is understating model uncertainty while KL-divergence  overstates it. Therefore, we construct new divergences which lie between polynomial and KL-divergence  in the sense that they are finite more often than the former and less often than the latter. Conversely, the associated worst cases are  infinite less often than under KL-divergence but more often than under polynomial divergence. The proposed divergence measures are for instance sensible choices for lognormal or Weibull reference models. A generalization of our approach from risk assessment to robust optimization, possibly in a dynamic setting, is a natural direction for future research.
\newpage
\renewcommand{\baselinestretch}{1.0} \small  \pagestyle{plain}
\begin{appendix}
	\section{Proofs}
	\subsection{Proofs of Section \ref{Ball}}

	{\sc{Proof of Proposition \ref{prop small balls}}.}\\
	For $M>0$, consider the probability distribution $\eta_M$ with density $g_M(x)=f(x)$ for $x \leq M$ and $g_M(x)=c(M)g(x)$ for $x>M$ where
	\[
	c(M)= \frac{\int_M^\infty f(x) dx}{\int_M^\infty g(x) dx}.
	\]
	By L'Hospital's rule, $c(M) \rightarrow 0$ for $M \rightarrow \infty$ since $f(M)/g(M) \rightarrow 0$ by assumption. In particular, there exists $M_0$ such that $c(M) \leq 1$ for all $M \geq M_0$. Since $F$ is continuous and strictly convex with $F(y)/y \rightarrow \infty$, there exists $y^*$ such that $F(y)$ is increasing for $y \geq y^*$ and that $F(y) \leq F(y^*)$ for $0 \leq y \leq y^*$. Moreover, since $g(x)/f(x) \rightarrow \infty$, there exists $x^*$ such that $g(x)/f(x) \geq y^*$ for all $x \geq x^*$. Thus, we have
	\[
	F\left( c \cdot\frac{g(x)}{f(x)}\right) \leq F\left( \frac{g(x)}{f(x)}\right)
	\]
	for all $c \in [0,1 ]$ and $x \geq x^*$. For $M \geq \max(x^*,M_0)$, we can thus bound the divergence between $\eta_M$ and $\nu$ through
	\[
	D_F(\eta_M|\nu) = \int_M^\infty F\left(c(M) \frac{g(x)}{f(x)}\right) f(x) dx \leq \int_M^\infty F\left(\frac{g(x)}{f(x)}\right) f(x) dx
	\]
	where we used that $F(1)=0$. Since $D_F(\eta|\nu) $ is finite, this upper bound on $D_F(\eta_M|\nu)$ becomes arbitrarily small for large $M$. Thus, we can find $M_\kappa \geq \max(x^*, M_0) $ with $D_F(\tilde{\eta}|\nu) \leq \kappa$ for the choice $\tilde{\eta}=\eta_{M_\kappa}$. The function $\tilde{\gamma}$ associated with $\tilde{\eta}$ is given by $\tilde{\gamma}(x) = \gamma(x) -\log(c(M_\kappa))$ for $x> M_{\kappa}$ and thus $\tilde{\gamma}(x)/ \gamma(x) \rightarrow 1$.\endpr

	{\sc{Proof of Proposition \ref{propKL}}.}\\
	
	The result follows  from the observation that
	\[
	D_{KL}(\eta|\nu)=\int_0^\infty \log\left(
	\frac{g(x)}{f(x)}
	\right)g(x) dx= \int_0^\infty \left(
	\varphi(x)-\gamma(x)
	\right)g(x) dx.
	\]\endpr

	{\sc{Proof of Proposition \ref{proppoly}}.}\\
	The proofs of both parts rely on the following claim: Suppose there exists $T \in \mathbb{R}$ such that
	\[
	\int_0^{\infty} f(x)^t dx= \infty
	\]
	for $t< T$ and
	\[
	\int_0^{\infty} f(x)^t dx< \infty
	\]
	for $ t>T$. Then  $D_{\alpha}(\eta|\nu)$ is finite if $\lim_{x \rightarrow \infty} h(x) > T$ and $D_{\alpha}(\eta|\nu)$ is infinite if $\lim_{x \rightarrow \infty} h(x) < T,$ where $h(x)=\alpha \gamma(x)/\varphi(x)-(\alpha-1)$.
	To prove the claim, it suffices to observe that finiteness of $D_{\alpha}(\eta|\nu)$ is equivalent to finiteness of
	\[
	\int_0^{\infty} \left(\frac{g(x)}{f(x)}\right)^\alpha f(x) dx = \int_0^{\infty} f(x)^{h(x)} dx
	\]
	and to recall the convergence of $h$ and the local boundedness properties of the integrand. For part (i) of the proposition, we thus need to show that the claim is applicable with
	$T=0$. It suffices to show that $f^t$ is integrable for any $t>0$, since $f^t$ and $f^{-t}$ cannot both have finite integrals over $[0, \infty)$.\footnote{
		To see this, observe that $1 \leq f^t(x)+f^{-t}(x)$ and $\int_0^\infty dx = \infty$.} To see that $f^t$ is integrable for any fixed $t>0$, recall that $f$ is bounded and that
	we can choose $K$ sufficiently large to ensure that $\varphi(x)/\log(x)> 2/ t$  for all $x >K$ since $\nu$ is of class (i) or (ii). This implies
	\[
	\int_0^\infty f(x)^t dx \leq \int_0^K f(x)^t dx + \int_K^\infty \frac{1}{x^2} dx < \infty.
	\]
	The argument for part (ii) of the proposition is similar. Here, $\nu$ is class (iii) distribution with
	$\varphi(x)/\log(x) \rightarrow c>1$. The claim can be applied with $T=\frac{1}{c}$ since,
	for large $x$, $f(x)^t$ behaves like $x^{-t \cdot c}$ which is integrable for $t\cdot c>1$, and not integrable otherwise.\endpr

	\subsection{Proofs of Section \ref{NewB}}
	
	{\sc{Proof of Lemma \ref{lem42}}.}\\
	$a$ and $b$ are simply chosen such that $F_{\Phi}$ and $F_{\Phi}'$ are continuous at the concatenation point.
	Since $F_{\Phi}'$  is continuous, $F_{\Phi}$ is convex as the two local definitions are convex. In particular, 
	convexity of $y \log(y)$ is clear. Convexity of $h(y)=ay \Phi^{-1}(\log(y))^\theta+b$ follows from  positivity of $a$ and positivity of
	\begin{eqnarray*}
		&&h''(y)= a \theta \Phi^{-1}(\log(y))^{\theta-2}/y  \nonumber \\
		&& \cdot
		\left( (\theta-1){\Phi^{-1}}'(\log(y))^2+
		\Phi^{-1}(\log(y))({\Phi^{-1}}'(\log(y))+{\Phi^{-1}}''(\log(y))  )
		\right)\nonumber
	\end{eqnarray*}
	for $y >\bar{y}$.
	Positivity of $h''$ follows from positivity, monotonicity and convexity of $\Phi^{-1}$ and $\theta \geq 1$.
	$F_\Phi(1)=0$ follows from $\bar{y} \geq 1$. $\lim_{y\rightarrow \infty} F_\Phi(y)/y =\infty$ follows from monotonicity of $\log$ and from the convexity and monotonicity of $\Phi^{-1}$.
	\endpr

	{\sc{Proof of Proposition \ref{ordering}}.}\\
	The case where $\eta$ is not absolutely continuous with respect to $\nu$ is clear. In the following we suppose that $\nu$ is a class (ii) distribution with associated functions $f, \varphi, \Phi$ and that $\eta$ has a density $g$. Assume first that $D_{F_\Phi}(\eta|\nu)<\infty$ . Since $\Phi$ is concave, we have $\Phi(z)\le z^\theta$ for $z$ large enough. This implies that there exists some $\tilde y$ such that $\log(y)\le \Phi^{-1}(\log(y))^\theta$ for all $y \ge \tilde y$.
	Therefore we have by assumption
	\begin{equation*}
	\begin{split}
	D^{KL}(\eta|\nu)&\le \int 1_{\{\frac{g(x)}{f(x)}\le \tilde y\}}\frac{g(x)}{f(x)}\log\left(\frac{g(x)}{f(x)}\right)f(x)dx\\
	&\quad+\int 1_{\{\frac{g(x)}{f(x)}> \tilde y\}}\frac{g(x)}{f(x)}\Phi^{-1}\left(\log\left(\frac{g(x)}{f(x)}\right)\right)^\theta f(x)dx<\infty.
	\end{split}
	\end{equation*}
	
	Next assume that $D^\alpha(\eta|\nu)<\infty$ for some $\alpha>1$. Since $\nu$ is of class (ii), it follows that $\lim_{x\to \infty}\frac{\Phi^{-1}(\varphi(x))}{\exp(\lambda \varphi(x))}=0$ for all $\lambda>0$. As $\lim_{x\to \infty} \varphi(x)=\infty$ this implies that $\frac{\Phi^{-1}(z)}{\exp(\lambda z)}\le 1$ for $z$ large enough. Now choosing $\lambda=(\alpha-1)/\theta$ we obtain $y^\alpha\ge y\Phi^{-1}(\log(y))^\theta$ for $y$ large enough. Following a similar argument as in the first part of the proof yields the claim.
	\endpr
	
	{\sc{Proof of Proposition \ref{suff_cond_finite_div}}.}\\
	By assumption there exists $\overline x>0$ such that $g(x)\le 1$ for all $x\ge \overline x$. Hence, there exist constants $K,\tilde K>0$ such that
	\begin{eqnarray*}
		D_{F_\Phi}(\eta|\nu)&=&\int _0^\infty F_\Phi\left(\frac{g(x)}{f(x)}\right)f(x)dx\le K+a\int_{0}^\infty1_{\{g(x)\ge \overline y f(x)\}}\Phi^{-1}\left(\log\left(\frac{g(x)}{f(x)}\right)\right)^\theta g(x)dx\\
		&\le& \tilde K + a\int_{\overline x}^\infty \Phi^{-1}(\varphi(x))^\theta g(x)dx.
	\end{eqnarray*}
	By construction we have $\Phi^{-1}(\varphi(x))\le Cx$ for $x$ large enough, which yields the claim.
	\endpr
	
	{\sc{Proof of Proposition \ref{prop44}}.}\\
	The proof proceeds in two major steps. We first provide a sufficient condition on $\Phi$ which implies the result. Then, we check this condition for the classes of nominal models we consider. The condition is as follows: Assume that for all constants $\alpha\ge 0$ and $s \in (0,1)$ it holds that
	\begin{equation}\label{cond_Phi}
	\liminf_{x\to \infty}\frac {\Phi^{-1}(\varphi(x)-\alpha\log(x))}{x^s}>0.
	\end{equation}
	Since $\nu$ is a class (ii) distribution and due to \eqref{cond_Phi} there exists a threshold $\overline x$ and constants $c,\delta>0$ such that $g(x)\ge \overline y f(x)$, $g(x)\ge \frac c{x^{t+1}}$ and
	\[
	\frac {\Phi^{-1}(\varphi(x)-(t+1)\log(x)+\log(c))}{x^s}\ge\delta.
	\]
	for all $x\ge \overline x$ and $s \in (0,1)$. Therefore, we have
	\begin{eqnarray*}
		D_{F_\Phi}(\eta|\nu)&\ge& K+a\int_{\overline x}^\infty \Phi^{-1}\left(\log\left(\frac{g(x)}{f(x)}\right)\right)^\theta g(x)dx\\ &\ge& K+a\int_{\overline x}^\infty \Phi^{-1}\left(\varphi(x)-(t+1)\log(x)+\log(c)\right)^\theta g(x)dx\\
		&\ge& K+a\delta^\theta \int_{\overline x}^\infty x^{s \theta} g(x)dx,
	\end{eqnarray*}
	for some constant $K$. Choosing $s=t/\theta \in (0,1)$ yields the claim, since $x^t g(x)\ge \frac cx$ for $x\ge \overline x$.
	
	It remains to check condition \eqref{cond_Phi} in the three cases. Assume first that $\Phi^{-1}$ is pseudo-regularly varying.
	Observe that for any $\alpha\ge 0$ we have $\frac{\varphi(x)-\alpha \log(x)}{\varphi(x)}\to 1$ as $x$ tends to $\infty$ since $\nu$ is a class (ii) distribution. It follows from Theorem 3.1 in \cite{buldygin2002properties} that $\Phi^{-1}$ preserves asymptotic equivalence of functions. It follows that $\frac {\Phi^{-1}(\varphi(x)-\alpha \log(x))}{\Phi^{-1}(\varphi(x))}\to 1$ as $x\to \infty$. By construction we have $\liminf \frac{\Phi^{-1}(\varphi(x))}{x}>0$ which implies \eqref{cond_Phi}.
	
	In the Weibull case $\Phi^{-1}$ is pseudo-regularly varying so (ii) follows from (i): We have $\Phi^{-1}(x)=\lambda x^{1/k}$, so $\frac {\Phi^{-1}(cx)}{\Phi^{-1}(x)}=c^{1/k}\to 1$ as $c\to 1$. Hence, $\Phi^{-1}$ is pseudo-regularly varying.

	In the generalized  lognormal case, $\Phi^{-1}$ is not pseudo-regularly varying so we verify \eqref{cond_Phi} directly.
	We have $\Phi^{-1}(x)=\exp(\sigma r^{1/r}x^{1/r})$. Set $x=\exp(1/h)$ and fix $\alpha>1$. Then we have
	\begin{align*}
	\log\left(\frac {\Phi^{-1}(\varphi(x)-\alpha\log(x))}{x^s}\right)&=\sigma r^{1/r}\left(\frac 1{r\sigma^r}|\frac 1h -\mu|^r+\frac 1h+\log(Z)-\frac \alpha h\right)^{1/r}-\frac {s}h \\
	&=\frac 1h \left(\vartheta(h)-s\right)
	\end{align*}
	with
	\[
	\vartheta(h)=\sigma r^{1/r}\left(\frac 1{r\sigma^r}|1-\mu h|^r+h^{r-1}+h^r\log(Z)-\alpha h^{r-1}\right)^{1/r}.
	\]
	Since $\vartheta(0)=1 >s$, $(\vartheta(h)-s)/h$ diverges to $+ \infty$ as $h \downarrow 0$ which implies \eqref{cond_Phi}.
	\endpr

	\subsection{Proofs of Section \ref{Appl}}
	
	To lighten the notation, we drop the subscript $\Phi$ from $F_{\Phi}$ throughout this section. In order to connect our claims to the results of \citet{breuer2013measuring}, we need to introduce some concepts from convex analysis. For $(\alpha_1, \alpha_2) \in \mathbb{R}^2$ we define
	\[
	K(\alpha_1,\alpha_2)=\int_0^\infty F^*(\alpha_1+\alpha_2x)f(x)dx,
	\]
	where $F^*(x)=\sup_{s\ge 0} (xs-F(s))$ is the convex conjugate of $F$.
	
	To prove Proposition \ref{prop51}, we first gather the following proposition from \citet{breuer2013measuring}'s Corollary 4.6, Lemma 4.1 and Theorem 4.2, using that under our smoothness assumptions we have $(F')^{-1}(x)=(F^*)'(x)$.

	\begin{proposition}\label{propA2}
		If $K(\alpha_1,\alpha_2)$ is finite and differentiable for all $(\alpha_1,\alpha_2) \in \mathbb{R}^2$,
		then the worst case in \eqref{wc} is finite and there exists $(\alpha_1^{wc},\alpha_2^{wc})\in \mathbb{R}\times \mathbb{R_+}$ such that the worst case model $\eta^{wc}$ has the density  $$g^{wc}(x)=(F')^{-1}(\alpha_1^{wc}+\alpha_2^{wc}x)f(x).$$ Moreover, $(\alpha_1^{wc},\alpha_2^{wc})\in \mathbb{R}\times \mathbb{R_+}$ are uniquely characterized by the conditions that $g^{wc}$ integrates to $1$ and that $D_{F}(\eta^{wc}|\nu)=\kappa$.
	\end{proposition}
	
	To complete the proof of Proposition \ref{prop51}, it thus suffices to check that the conditions in that proposition imply the conditions on $K(\alpha_1,\alpha_2)$ in Proposition \ref{propA2}.\footnote{The conditions on $K(\alpha_1,\alpha_2)$ of Proposition \ref{propA2} imply that $K$ fulfills a property called essential smoothness and that its effective domain, i.e., the subset of $\mathbb{R}^2$ where $K$ is finite, is the whole space. If necessary, one can easily extend the result to the case where the effective domain is an open subset of $\mathbb{R}^2$ which contains some points with $\alpha_2>0$. See footnote 7 in \citet{breuer2013measuring}.} This is the content of the next lemma:
	
	\begin{lemma}\label{lemA3}
		Let $\nu$ be a class (ii) distribution with associated functions $f$, $\varphi$, $\Phi$ and fix $\theta>1$.
		If for all $\alpha_1 \in \mathbb{R}$ and $\alpha_2 >0$ there exists $c>0$ such that $I(c)$ defined in \eqref{condIc} is well-defined and finite, then $K(\alpha_1,\alpha_2)$ is finite and differentiable for all $(\alpha_1,\alpha_2) \in \mathbb{R}^2$.
	\end{lemma}
	\proof
	We show that $K(\alpha_1, \alpha_2)$ is finite and that the terms
	\[
	K_1(\alpha_1, \alpha_2)= \int_0^\infty {F^*}'(\alpha_1+\alpha_2x)f(x)dx,
	\]
	and
	\[
	K_2(\alpha_1, \alpha_2)= \int_0^\infty x {F^*}'(\alpha_1+\alpha_2x)f(x)dx =\int_0^\infty x {F'}^{-1}(\alpha_1+\alpha_2x)f(x)dx
	\]
	which are proportional to the $\alpha_1$- and $\alpha_2$-derivatives of $K$ are finite.
	Differentiability then follows by an application of Lebesgue's dominated convergence theorem. Observe that $F^*$ is bounded on every interval $(-\infty,a]$, $a\in \mathbb R$. This implies that for $\alpha_2\le 0$ we have that $K(\alpha_1,\alpha_2)<\infty$ for all $\alpha_1\in \mathbb R$. Since ${F^*}'$ is increasing we also obtain finiteness of $K_1$ and $K_2$ in this case. Next, fix $(\alpha_1,\alpha_2)\in \mathbb{R}\times \mathbb{R_+}$. By construction $F':(0,\infty) \to (-\infty,\infty)$ is a bijection. This implies for $x\in \mathbb R$ that $x {F^*}'(x)= F^*(x)+F((F')^{-1}(x))$. By the monotonicity of ${F^*}'$ and since $F$ is bounded from below, it follows that finiteness of $K_2$ implies finiteness of all three integrals. Performing a change of variables $x=(F'(e^y)-\alpha_1)/\alpha_2$, there exist constants $L$ and $c>0$ so that we can write $K_2$ as
	\[
	K_2(\alpha_1,\alpha_2)=L +\int_c^\infty e^y F'(e^y)F''(e^y)e^y f\left(\frac{F'(e^y)-\alpha_1}{\alpha_2}\right) dy
	\]
	For $x$ large enough $F$ is of the form $F(x)=x\psi(\log(x))+b$ with $\psi(x)=a\Phi^{-1}(x)^\theta$. Observe that $F'(e^x)=\psi(x)+\psi'(x)$ and $F''(e^x)e^x =\psi'(x)+\psi''(x)$ so that we obtain
	\[
	K_2(\alpha_1,\alpha_2)=L +\int_c^\infty e^y (\psi(y) + \psi'(y))(\psi'(y)+\psi''(y)) f\left(\frac{F'(e^y)-\alpha_1}{\alpha_2}\right) dy
	\]
	for sufficiently large $c$. Since the last integral is $I(c)$, finiteness of $I(c)$ concludes the proof.
	\endpr
	
	We have thus proved Proposition \ref{prop51} and turn to Corollary \ref{cor52} and Proposition \ref{prop53}.
	
	{\sc{Proof of Corollary \ref{cor52}}.}\\
	In the Weibull case, plugging the definitions of $f$ and $\psi(x)=a\lambda^\theta x^\frac \theta k$ into $I(c)$, we obtain
	\[
	C \int_c^\infty (y^{\beta}+ \beta y^{\beta-1})(y^{\beta-1}+(\beta-1)y^{\beta-2})(a\lambda^\theta(y^\beta+\beta y^{\beta-1})-\alpha_1)^{k-1}e^y
	e^{-\left(\frac{a\lambda^\theta(y^\beta+\beta y^{\beta-1})-\alpha_1}{\lambda\alpha_2}\right)^k}dy
	\]
	with $\beta=\theta/k$ and $C>0$. The leading part of the last exponential is of order $-(a\lambda^{\theta-1}/\alpha_2)^k y^\theta$. Since $\theta>1$, the integral is thus finite.
	
	In the generalized lognormal case, we have
	$\psi(x)=a\Phi^{-1}(x)^\theta=ae^{  \sigma \theta (rx)^{1/r}}$ and thus
	\begin{eqnarray*}
		\psi'(x)&=& \psi(x) \sigma \theta (r x)^{\frac{1}{r}-1} \\
		\psi''(x)&=& \psi(x)  \sigma \theta (rx)^{\frac{2}{r}-2}  \left(\sigma \theta +(1-r)(r x)^{-\frac1{r}}  \right).
	\end{eqnarray*}
	Using $F'(e^y)=\psi(y)+\psi'(y)$ we obtain
	\[
	I(c) = \int_{c}^\infty e^y H(y) \exp \left(- \frac{1}{r \sigma^r}|\log(\psi(y)+ \psi'(y)-\alpha_1) +\log(\alpha_2)  - \mu) |^r \right) dy
	\]
	where the term
	\[
	H(y)= \frac{\alpha_2 ( \psi(y)+ \psi'(y))(\psi'(y)+\psi''(y))}{(\psi(y)+\psi'(y)-\alpha_1)Z}
	\]
	is of leading exponential order $m \cdot x^{\frac1r}$ for some constant $m$. Moreover, $H$ is positive for sufficiently
	large $y$ since $\psi$ gets arbitrarily large and $\psi''$ is positive for sufficiently large $y$.
	We can write
	\[
	I(c) = \int_{c}^\infty e^y H(y) e^{- \frac{1}{r \sigma^r}|\log\left( \frac{\psi(y)}{a}\right) + G(y)|^r } dy.
	\]
	with
	\[
	G(y) = \log\left( \frac{a(\psi(y) + \psi'(y)-\alpha_1) }{ \alpha_2 \psi(y)}\right)  - \mu.
	\]
	$G(y)$ is bounded since $\psi$ goes to $+ \infty$ and $\psi'/\psi$ to zero.
	Applying the elementary inequality
	\[
	-|x_1- x_2|^r \leq -\alpha(\varepsilon) |x_1 |^r + \beta(\varepsilon) |x_2|^r, \;\;\;  \alpha(\varepsilon):= (1- \varepsilon)^{r-1}, \;\;
	\beta(\varepsilon):= \frac{(1- \varepsilon)^{r-1}}{\varepsilon^{r-1}},
	\]
	which holds for any $x_1,x_2 \in \mathbb{R}$, $r \geq 1$ and $\varepsilon \in (0,1)$, we obtain the estimate
	\[
	I(c) \leq \int_{c}^\infty e^y H(y) e^{- \frac{\alpha(\epsilon)}{r \sigma^r}|\log\left( \frac{\psi(y)}{a}\right)|^r} e^{\frac{\beta(\varepsilon)}{r \sigma^r}|G(y)|^r  }  dy.
	\]
	The factor involving $G$ is bounded for any $\varepsilon$, so finiteness of $I(c)$ follows from
	\[
	e^{- \frac{\alpha(\varepsilon)}{r \sigma^r}|\log\left( \frac{\psi(y)}{a}\right)|^r} = e^{- \theta \alpha(\varepsilon) y}
	\]
	and $\theta \alpha(\varepsilon) >1$ for sufficiently small $\varepsilon$.
	\endpr
	
	{\sc{Proof of Proposition \ref{prop53}}.}\\
	For the Weibull case, observe that for $y\ge \bar y$ we have $F'(y)=a\lambda \log(y)^{\frac{\theta}{k}}\left(1+\frac {\theta \lambda}{k \log(y)}\right)$. Therefore $F'$ is pseudo-regularly varying (see Footnote \ref{def_prv} for the definition). In particular, it follows from Theorem 3.1 in \cite{buldygin2002properties} that $F'$ preserves asymptotic equivalence of functions. Hence, it suffices to show that $\alpha_1+\alpha_2x$ is asymptotically equivalent to $F'\left(\frac{\overline{g}(x)}{f(x)}\right)$. We have
	\[
	F'\left(\frac{\overline{g}(x)}{f(x)}\right)=a\lambda\left(\log(\overline{g}(x)/f(x))\right)^{\frac \theta k}\left(1+\frac {\theta \lambda}{k\left(\log(\overline{g}(x)/f(x))\right)}\right)
	\]
	By our choice of $\overline{g}$, we have
	\[
	\lim_{x\to \infty}\frac{a\lambda\left(\log(\overline{g}(x)/f(x))\right)^{\frac \theta k}}{\alpha_1+\alpha_2x}=1,
	\]
	which implies that $g$ is asymptotically equivalent to $\overline{g}$. For the lognormal case,  we have $$F'(y)=a\exp\left(\theta\sqrt{2\sigma^2\log(y)}\right)\left(1+\theta\sqrt{\frac {\sigma^2}{2\log(y)}}\right)$$ for  $y\ge \bar y$. We next verify that $F'$ is pseudo-regularly varying so that it preserves asymptotic equivalence of functions. Indeed, it follows from
	\[
	\sqrt{\log{cy}}-\sqrt{\log{y}}=\frac {\log(c)}{\sqrt{\log{cy}}+\sqrt{\log{y}}}\to 0
	\]
	as $y\to \infty$, that $\lim_{y\to \infty}\frac {F'(cy)}{F'(y)}=1$. It remains to show that $\alpha_1+\alpha_2x$ is asymptotically equivalent to $F'\left(\frac{\overline{g}(x)}{f(x)}\right)$. We have
	\[
	\log\left(\frac{\overline{g}(x)}{f(x)}\right)=\frac 1{2\sigma^2 \theta^2}\log\left(\frac {\alpha_1+\alpha_2x}{a}\right)^2
	\]
	It follows that
	\[
	F'\left(\frac{\overline{g}(x)}{f(x)}\right)=(\alpha_1+\alpha_2x)\left(1+\frac {\theta^2\sigma^2}{\log\left(\frac 1a(\alpha_1+\alpha_2x)\right)}\right),
	\]
	which implies that $F'\left(\frac{\overline{g}(x)}{f(x)}\right)$ is asymptotically equivalent to $\alpha_1+\alpha_2x$.
	\endpr

\end{appendix}

\end{document}